\documentclass{aa}
\usepackage{txfonts}
\usepackage{graphicx}
\usepackage{natbib}
\usepackage{multirow} 
\bibpunct{(}{)}{;}{a}{}{,} 

\newcommand{\micron}{\mbox{$\mu$m}}

\begin{document}

	\title{Atomic jet from SMM1 (FIRS1) in Serpens uncovers non-coeval binary companion}
	\author{O.~Dionatos\inst{1,2,3}
          \and
	  J.~K.~J{\o}rgensen
	   \inst{3,2}
	  \and
	   P.~S.~Teixeira\inst{1}
	   \and
	   M.~G{\"u}del\inst{1}
	   \and
	  E.~Bergin
	   \inst{4} 
          }

\institute{
		University of Vienna, Department of Astrophysics,  T{\"u}rkenschanzstrasse 17,
A-1180, Vienna, Austria\\
		\email{odysseas.dionatos@univie.ac.at} \\
	\and 
		Centre for Star and Planet Formation, Natural History Museum of
Denmark, University of Copenhagen,  {\O}ster Voldgade 5 -- 7, DK-1350 Copenhagen K.
Denmark\\
         \and
		Niels Bohr Institute, University of
Copenhagen. Juliane Maries Vej 30, DK-2100 Copenhagen {\O}. Denmark\\\
     	 \and
	Department of Astronomy, University of Michigan, 500 Church Street, Ann Arbor, Michigan 48109, USA\\
             }

\abstract
{We report on the detection of an atomic jet associated with the protostellar source SMM1 (FIRS1) in Serpens.  The morphological and physical characteristics of the atomic jet suggest the existence of a more evolved protostellar companion to the Class~0 source SMM1.}
{To disentangle the molecular and atomic emission around the protostellar source Serpens-SMM1, identify the emission line origin and assess the evolutionary stage of the driving sources.}
{The surroundings of SMM1 were mapped with the \textit{Spitzer} Infrared Spectrograph (IRS) in slit-scan mode. The complex outflow morphology of the molecular (H$_2$) and atomic ([FeII], [NeII], [SiII], [SI]) emission from \textit{Spitzer} is examined along with deconvolved \textit{Spitzer} IRAC and MIPS images and high-velocity CO $J$=3-2 outflow maps. The physical conditions of the atomic jet are assessed assuming LTE, non-LTE conditions, and shock models.}
{The atomic jet is firmly detected in five different [FeII] and [NeII] lines, with possible contributions from [SI] and [SiII]. It is traced very close to SMM1 and peaks at $\sim 5\arcsec$ from the source at a position angle of $\sim 125^\circ$. H$_2$ emission becomes prominent at  distances $> 5\arcsec$ from SMM1 and extends at a position angle of 160$^\circ$. The morphological differences suggest that the atomic emission arises from a companion source, lying in the foreground of the envelope surrounding the embedded protostar SMM1. The molecular and atomic emission  disentangle the large scale CO emission into two distinct bipolar outflows, giving further support to a proto-binary source. Analysis at the peaks of the [FeII] jet show that emission arises from warm and dense gas (T $\sim$~1000~K, n$_e$~$\sim$10$^5$ - 10$^6$~cm$^{-3}$). The mass flux of the jet derived independently for the [FeII] and [NeII] emission is $\sim10^7$ M$_{\odot}$~yr$^{-1}$, pointing to a more evolved Class~I/II protostar as the driving source. Comparisons of the large-scale  outflow and atomic jet momentum fluxes show that the latter has adequate thrust to support the CO outflow.}
{The atomic jet detected by \textit{Spitzer} gives for the first time the opportunity to disentangle the complex outflow morphology around SMM1 into two precessing outflows. The morphological and physical properties of the outflows reveal that SMM1 is a non-coeval proto-binary source. The momentum flux of the atomic jet indicates that the companion to the deeply embedded Class~0 protostar SMM1 is a more evolved Class~I/II source.}


\maketitle


\section {Introduction}
\label{sec:1}

The observational manifestations of protostellar ejecta show morphological trends which follow the evolutionary stage of their parent bodies. In  young embedded Class~0 protostars, interferometric observations of fast moving gas reveal highly collimated ``molecular'' jets \citep[e.g. HH211,][]{Gueth:99a}. Large scale outflows in embedded (Class~0/I) protostars reflect the interactions between ejecta and the surrounding dense medium. These interactions result in the gradual dispersion of the envelope \citep{Arce:06a}, and eventually the protostellar ejecta are observed in the infrared and visual wavelengths as ``atomic'' jets. Morphological characteristics,  such as the high degree of collimation and knotty structure observed in both ``molecular'' and ``atomic'' jets are suggestive of a common formation and collimation mechanism \citep{Cabrit:07a}. Given that the ejecta in the comparatively more evolved Class~II sources are arguably atomic  \citep[traced down to a few AU from the protostar, e.g.,][]{Agra-Amboage:11a}, obscured atomic jets may also be responsible for their molecular counterparts  seen in embedded protostars. Indications of the existence of such atomic jets from embedded sources have been found in few cases \citep[e.g., L1448-mm,][]{Dionatos:09a}. However, mid-infrared surveys of Class~0 protostars \citep{Lahuis:10a} have detected extended [FeII] and [NeII] emission that could be attributed to jets in only $\sim$10\% of the cases. Atomic lines in the far infrared ([OI] and [CII]), commonly traced around embedded sources \citep[e.g.,][]{Green:13a}, may be excited by different mechanisms so they cannot be used as safe indicators of atomic jets  \citep{Visser:12a}. It therefore remains unclear whether molecular jets are manifestations of underlying, obscured atomic jets or represent direct ejecta from embedded protostars \citep{Panoglou:12a}.        

In addition to their morphological characteristics, protostellar ejecta provide insights on the mass accretion/ejection processes that control protostellar evolution. For embedded sources, it has been shown that the momentum flux, or ``thrust'', of outflows strongly correlates with the luminosity of a protostar which is directly linked to accretion \citep{Bontemps:96a}. Analogous accretion/ejection correlations have been shown to hold also for T~Tauri stars  \citep{Hartigan:95a}. Furthermore, such correlations are found to be valid for both low and high mass protostars, providing evidence for a common formation mechanism independently of the protostellar mass \citep{Wu:04a, Zhang:05a}, and give strong evidence that the accretion rate drops as a protostar evolves. The exact link between accretion and ejection is still debated, however most theoretical models \citep[e.g. X-wind and disk-wind models,][respectivelly]{Shu:94a, Ferreira:97a} agree that there is a strong link between the momentum output of the jet/wind and the accretion rate. Therefore the jet/outflow mass loss rate and momentum flux can provide indirect measures of the mass accretion rate, and thus constrain the evolutionary stage of a protostar.         

Among nearby embedded protostars, SMM1 in Serpens is an outstanding source that has drawn lots of attention over the last 30 years. It is the most prominent source in Serpens \citep[$d=260 - 415$ pc;][respectively, the former value adopted in this work]{Straizys:03a, Dzib:10a} associated with energetic mass ejection phenomena.  Strong, extended outflows traced in a number of molecular and atomic tracers \citep[e.g.,][]{Goicoechea:12a, Davis:99a, Testi:98a} co-exist with strong radio and maser emission \citep[e.g.,][respectively]{Curiel:93a, vanKempen:09a}. The complex morphology of the outflows seen at interferometric resolutions \citep{Hogerheijde:99a,Testi:98a} has been proposed to be driven by a proto-binary system \citep{White:95a, Dionatos:10b}. 

Continuum observations of SMM1 have been the subject of intense scrutiny.  Millimeter and sub-millimeter parts of the spectral energy distribution (SED) reveal an extended envelope and a compact peak corresponding to masses of 8.7 M$_{\odot}$ and 0.9 M$_{\odot}$, respectively \citep{Hogerheijde:99a}. The suggested $M_{env}/M_{star} < 0.5$ corresponds to a young, deeply embedded Class~0 protostar. Still, the mid-IR part of the spectrum shows significant IRAC emission \citep{Winston:07a} which has been hard to attribute to a Class~0 source.  Fitting the full SED, \citet{Davis:99a} reported a bolometric temperature of $\sim$ 38 K, and assigned SMM1 an intermediate evolutionary stage between Classes~0 and I. More recently, \citet{Enoch:09a} modeled SMM1 as a luminous Class~0 source and attributed the near and mid-IR emission to a massive disk of $\sim$1~M$_{\odot}$. \citet{Choi:09a}, using mm and cm continuum interferometric observations, reported on the existence of a weaker secondary continuum source positioned at  $\sim$~1.8$\arcsec$ (500 AU) to the NW, suggesting that SMM1 is a protostellar binary. From the spectral index between 6.9~mm and 7~cm, they argued that the secondary peak is a disk source that dominates the IR part of the SED. This was previously suggested based on the mismatch between the positions of the near-IR source EC41 and the millimeter continuum source SMM1 \citep{Eiroa:89a, Hodapp:99a, Eiroa:05a}. It should be noted that EC41 and SMM1-b are not spatially associated. The possible companion observed in \citet{Choi:09a} is also detected in 230 GHz interferometric continuum maps of \citet{Enoch:09a}, however it was interpreted as a condensation along the outflow.  

In this paper we employ \textit{Spitzer} spectro-imaging observations to disentangle the emission around SMM1. We describe the data reduction in Sect.~\ref{sec:2} and discuss the spatial morphology of atomic and molecular (H$_2$ and CO) emission lines   in Sect.~\ref{sec:3}.  In Sect.~\ref{sec:4} we derive the excitation conditions and dynamics of the atomic jet, and constrain its possible progenitor. Our conclusions are summarized in Sect.~\ref{sec:5}.

\section{Data Reduction}\label{sec:2}

The area around SMM1 \citep[$\alpha_{J2000}$=18$^h$29$^m$49$^s$.8,  $\delta_{J2000}$=+01$^d$15$^m$20$^m$.6,][]{Choi:09a} was observed on June 5, 2009 as part of the ``Searching for the Missing Sulfur in the Dense ISM'' program (E. Bergin, P.I.). The short-high (SH) and long high (LH) modules of the \textit{Spitzer Infrared Spectrograph} \citep[IRS,][]{Houck:04a} were employed, providing a wavelength coverage between 10 and  37~$\micron$ at a resolution of R~$\sim$~600. 
Observations were performed in slit-scan mode consisting of consecutive
integrations after shifting the slit to the parallel and perpendicular
directions relative to the slit length, until the desired area is covered.
The SH and LH scans consist of grids of 6~$\times$~21 and 11~$\times$~10 observations, respectively, centered at $\alpha_{J2000}$=
18$^h$29$^m$48$^s$.7, $\delta_{J2000}$=+01$^d$15$^m$10$^m$.4. The integration time per pointing is  30 and 6 seconds for the SH and LH modules, respectively.

Spectra were retrieved from the \textit{Spitzer Heritage Archive} (SHA). Initial data processing was performed with  version S18.7 of the
\textit{Spitzer} Science Center pipeline. Spectral data cubes were compiled
using the CUBISM software \citep{Smith:07a}.  Bad/rogue pixels were removed using dedicated off-target observations. 
Emission line maps were reconstructed  through customized procedures. In these, for each spatial pixel (or \textit{spaxel}) of a
data-cube, the flux for each spectral line of interest was calculated by
fitting a Gaussian after subtracting a local first or second order polynomial
baseline. The resulting line intensity maps for the IRS data have a square
spaxel of side equal to the width of the low resolution IRS modules
(4.7$\arcsec$ and 11.1$\arcsec$ for the SH and LH, respectively), while
the instrumental point spread function of \textit{Spitzer} ranges between $\sim 3\arcsec$ at 10 $\mu$m to
11$\arcsec$ at 38 $\mu$m. 

The wavelength range covered by the SH and LH modules is limited towards the short-end to 10~$\micron$. As a result, the higher energy rotational H$_2$ transitions (S(3) - S(7)) and other possible atomic lines are not included in the current data. However, lower energy H$_2$ and atomic lines, mainly from [FeII] but also from [NeII], [SI] and [SiII] are detected (Table~\ref{tab:1}), indicative of a very energetic environment.  

\begin{figure}
\centering
\includegraphics[width=0.45\textwidth]{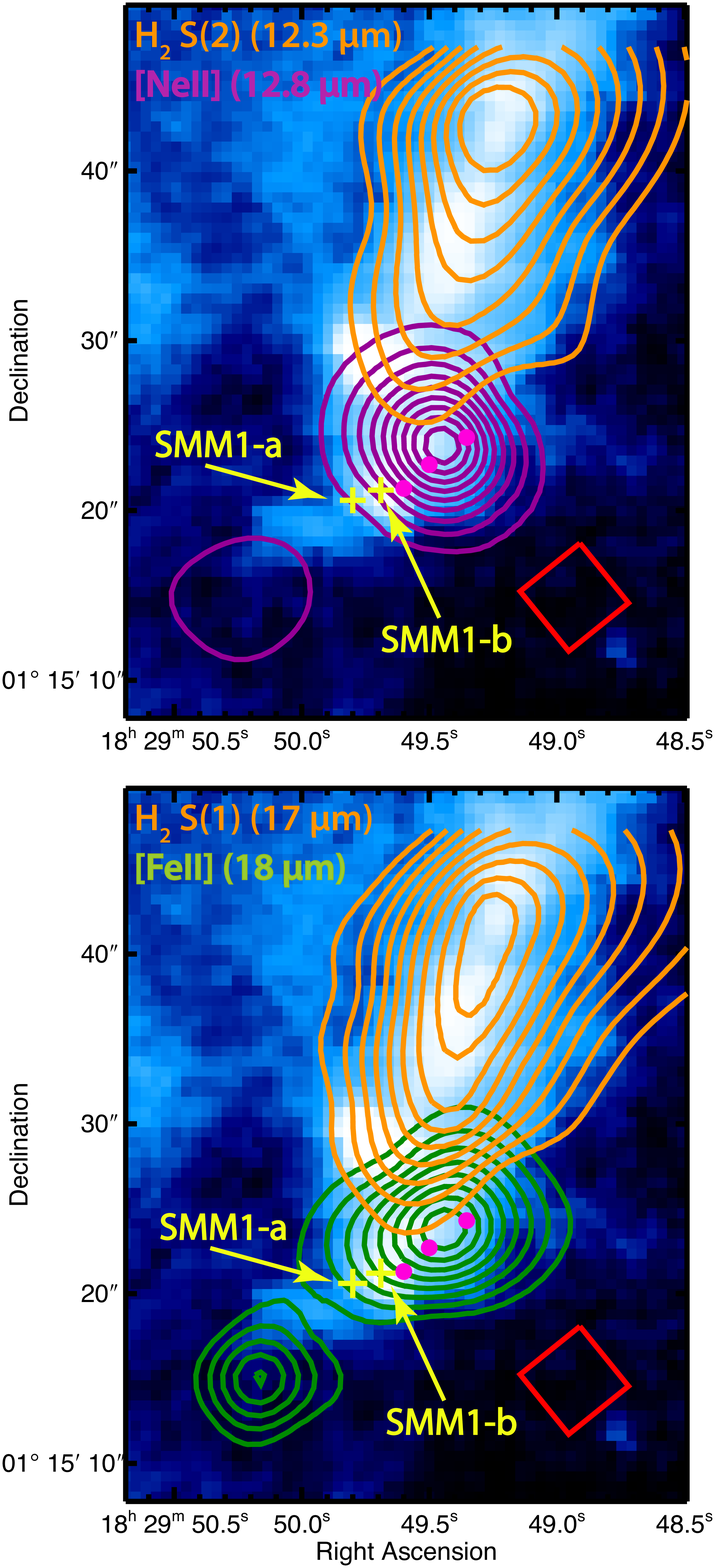}
\caption{Emission line maps of the [NeII] $^2$P$_{\frac{1}{2}}$ -- $^2$P$_{\frac{3}{2}}$  and H$_2$ S(2) lines (top panel) and the [Fe II] $^4$F$_{\frac{7}{2}}$ -- $^4$F$_{\frac{9}{2}}$ and  H$_2$ S(1) lines  (lower panel). On both panels line emission is superimposed on an IRAC band-2 (4.5~$\micron$) image.  The positions of SMM1-a and SMM1-b (to the SE and NW, in respect to each other) from \citet{Choi:09a} are marked with crosses, and the SH spaxel size and orientation is marked as a (red) square in the lower right corner of each panel. The (magenta) filled circles show the positions of the 163 GHz H$_2$O maser emission \citep{vanKempen:09a}, coincident with the NW lobe of the atomic emission. Contour levels are  at 10\% -- 90\% of the peak emission for each transition, as listed in Table~\ref{tab:1}. Each pair of H$_2$ and atomic lines on both panels is separated by less than 1 $\micron$ and therefore should be equally affected by reddening. However, the atomic emission is traced very close to the base of the outflows whereas the H$_2$ emission becomes significant only at angular distances $> 5 \arcsec$ (corresponding to $\sim$ 1300 AU for the adopted distance).} 
\label{fig:1}
\end{figure}

\section{Outflow morphology}\label{sec:3}
 
The upper panel of Fig.~\ref{fig:1} presents the spatial distribution of the [NeII] $^2$P$_{\frac{1}{2}}$ -- $^2$P$_{\frac{3}{2}}$ transition at 12.8~$\micron$  along with the observed emission distribution from the H$_2$ S(2) line centered at 12.3 $\micron$. The lower panel of the same figure shows the emission pattern from [FeII] $^4$F$_{\frac{7}{2}}$ -- $^4$F$_{\frac{9}{2}}$ and H$_2$ S(1) lines, centered at 18 and 17~$\micron$, respectively. All lines detected with the SH module are superimposed on an IRAC band-2 image, centered at 4.5~$\micron$. The positions of 6.9~cm continuum sources SMM1-a (to the SE) and SMM1-b (to the NW) from \citet{Choi:09a}, along with the locations of 183~GHz H$_2$O maser hotspots from \citet{vanKempen:09a}, are also indicated. 

\begin{figure}
\centering\
\resizebox{\hsize}{!}{\includegraphics{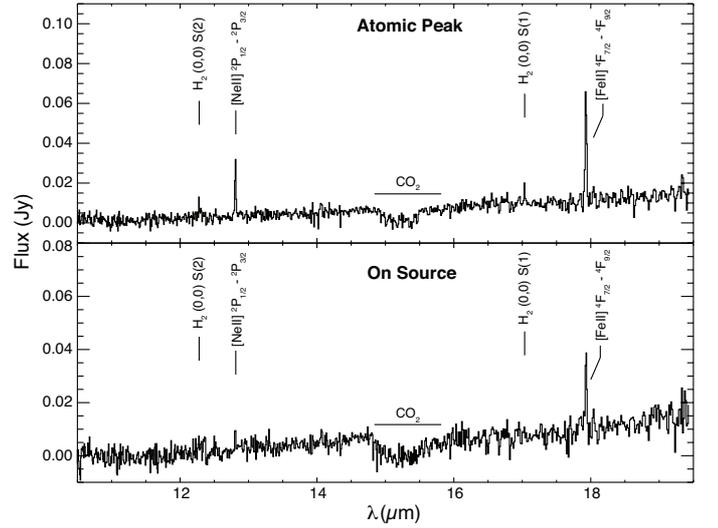}}
\caption{IRS spectra from the SH module (10-20~$\micron$) corresponding to the spaxels encompassing the on-source and atomic peak positions in Fig.~\ref{fig:1} (lower and upper panels, respectively). Both positions are dominated by strong atomic emission lines. No H$_2$ emission is detected at the on-source spaxel. CO$_2$ ice absorption bands are prominent in both spectra, indicating extended continuum emission to angular distances $>$~5~$\arcsec$ and a foreground absorbing layer.}
\label{fig:2}
\end{figure}

The H$_2$ line maps show very similar morphologies following the bright ridge recorded in the IRAC image at a position angle of $\sim$ 340$^\circ$. H$_2$ emission becomes apparent at angular distances $>$~$5\arcsec$ and peaks at $\sim$~$20\arcsec$ from the position of SMM1-a. The [NeII] and [FeII] lines are detected on the positions of SMM1-a and SMM1-b, and exhibit two symmetric lobes extending to the NW and SE, peaking at distances of $\sim$~$5\arcsec$. The strongest peak to the NW from the centimeter sources is aligned with the positions of H$_2$O maser detections. The atomic emission extends along a  position angle of 125$^\circ$,  which is in roughly the same direction but still different from the one traced by H$_2$. Figure~\ref{fig:2} presents IRS/SH spectra in the range between 10~$\micron$ and 20~$\micron$ for the spaxels pertaining to the source and atomic emission maxima positions (lower and upper panels, respectively). At the on-source position, the [FeII] and [NeII] lines are prominent, however no H$_2$ is detected. At the position where the atomic lines peak, H$_2$ lines are detected but they are weaker than the atomic ones. The CO$_2$ ice absorption band at $\sim$15~$\micron$ is prominent in both positions. This band appears superimposed on continuum suggesting that the continuum emission extends up to $\sim$5~$\arcsec$ or more from SMM1. The CO$_2$ band likely originates from ice mantles on dust grains, located in a foreground layer (see also Sect.~\ref{sec:4}).  

\begin{figure*}
\centering\
\resizebox{\hsize}{!}{\includegraphics{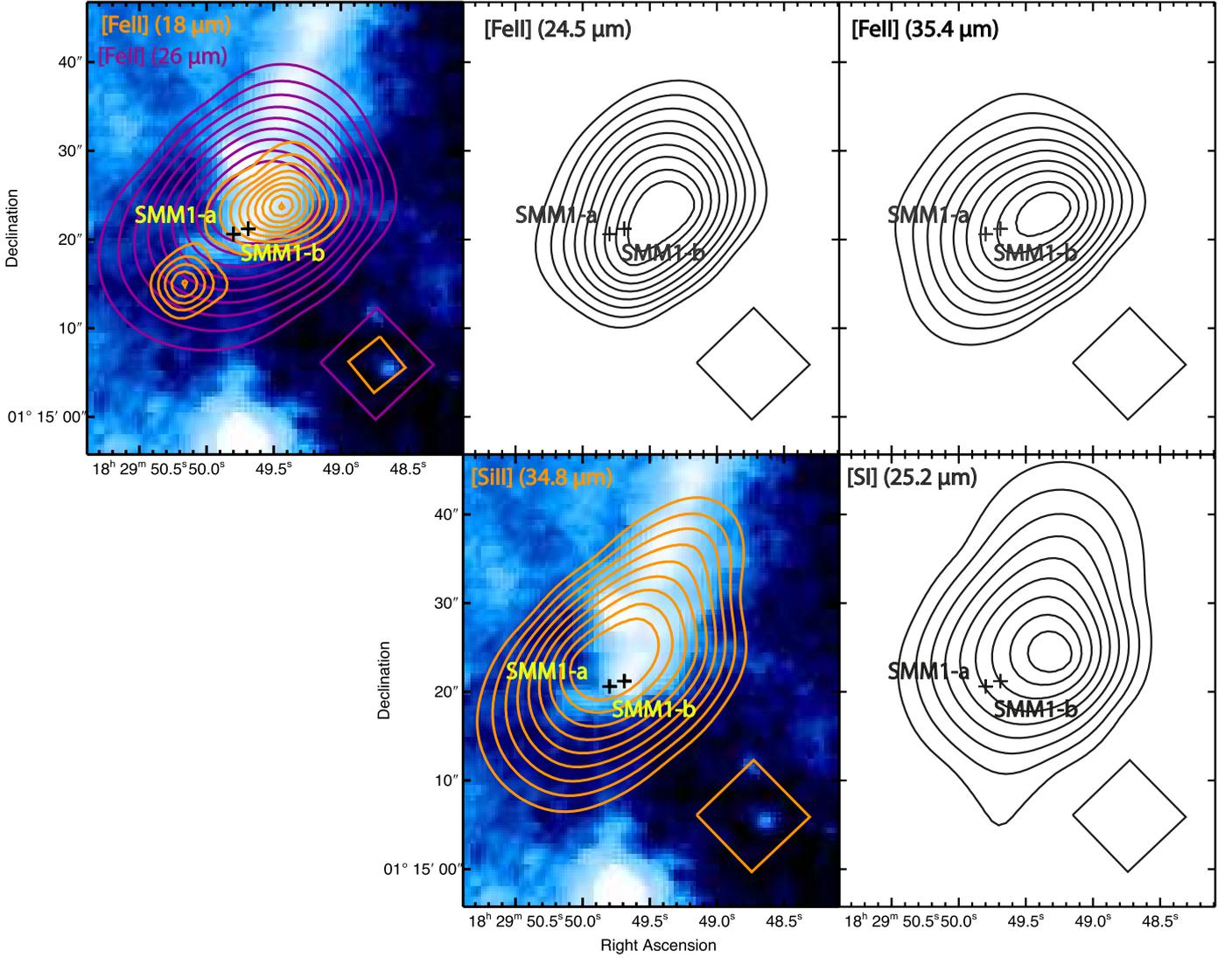}}
\caption{Spectral maps of atomic lines detected with the LH module: upper panels show the distribution of [FeII] lines and lower panels the morphology of [SiII] and [SI]. The contour maps on the leftmost panels are superimposed on an IRAC band-2 image. The LH and SH module spaxel size and orientation is presented at the lower right corner of each panel. The [FeII] 18 $\micron$ morphology mapped with the SH module (Fig. \ref{fig:1}) is presented as reference on the upper left panel along with the 26 $\micron$ [FeII] line. Despite the fact that the LH footprint covers by a factor of 4 more area than the SH one, [FeII] lines show a consistent morphology. In contrast, [SiII] and [SI] follow closer the H$_2$ emission pattern, also traced by the bright ridge in the IRAC image. Contour levels are  at 10\% -- 90\% of the peak emission for each transition, listed in Table~\ref{tab:1}}
\label{fig:3}
\end{figure*}

In Fig.~\ref{fig:3}, we present the emission pattern of the atomic lines detected with the LH module. The upper panels show the [FeII] transitions at 26, 24.5 and 35.4~$\micron$. The emission pattern of all [FeII] lines shows a very similar morphology to the pattern traced by the [FeII] 18~$\micron$, despite the inferior angular resolution of the LH compared to the SH module. [FeII] lines are detected very close to the continuum sources and a strong peak towards the NW at $\sim 5\arcsec$ from SMM1 is detected in all cases. The weaker peak at the SE is not resolved in the 26 $\micron$ and 35.4 $\micron$  maps, which however preserve the overall morphology of the 18 $\micron$ [FeII] line. The 24.5 $\micron$ [FeII] emission is not resolved at the SE peak. The [SI] and [SiII] lines presented in the lower panels of Fig.~\ref{fig:3} peak around the same location as the [FeII] lines, but also show elongated structures to the NW with a closer resemblance to the H$_2$.

\begin{figure*}
\centering\
\resizebox{\hsize}{!}{\includegraphics{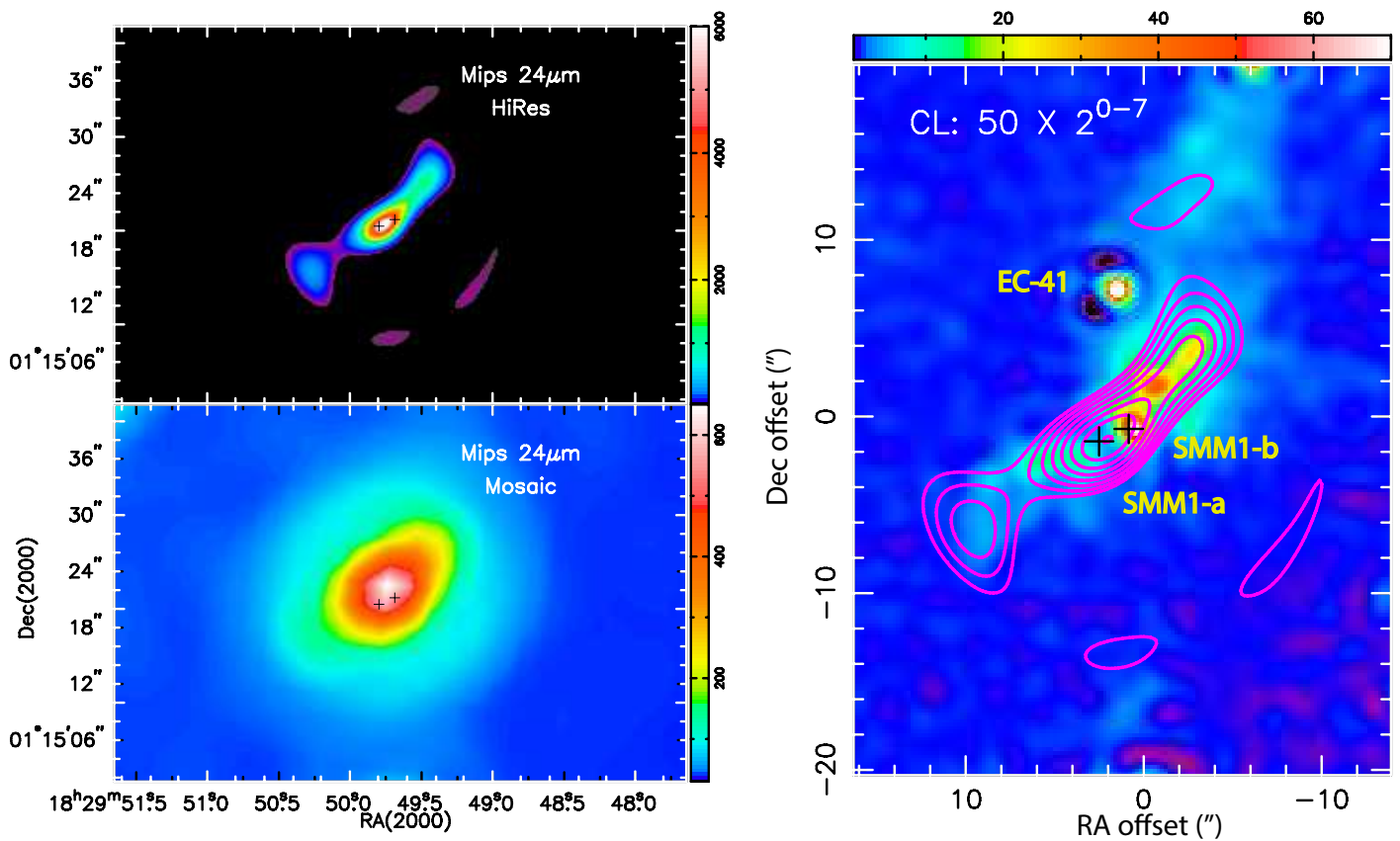}}
\caption{\textit{Left}: MIPS 24 $\mu$m mosaic image before and after HiRes deconvolution (lower and upper panels, respectively). The morphology delineated in the mosaic image at a resolution of 5~$\arcsec$ is in excellent agreement with the LH module spectral line maps shown in Fig.~\ref{fig:3}. The HiRes MIPS image reaching a resolution of 2$\arcsec$, compares directly to the atomic jet traced in the SH line maps (Fig.~\ref{fig:1}).  \textit{Right}:  MIPS HiRes image (purple contours) superimposed on an IRAC~4.5~$\mu$m HiRes image. The IRAC image is dominated by highly excited H$_2$ emission and is in good agreement with the morphology of the atomic jet seen in the MIPS HiRes band.  The companion source SMM1-b is coincident with a bright spot of emission in the IRAC channel, while the Class 0 source SMM1-a shows no association with the emission in the IRAC band. Images provided by T. Velusamy, private communication \citep{Velusamy:14a}.}
\label{fig:3a}
\end{figure*}

The extended emission seen in the Spitzer/MIPS Mosaic image at 24~$\micron$  (Fig.~\ref{fig:3a}) is dominated by the [SI] and [FeII] lines encompassed within the band \citep{Velusamy:07a, Velusamy:11a}. At an angular resolution of 5~$\arcsec$ the Mosaic image is in excellent agreement with the line maps morphologies presented in Fig.~\ref{fig:3}.   Deconvolution methods  such as the HiRes technique \citep[e.g.][]{Velusamy:08a, Velusamy:14a} can enhance the angular resolution of Spitzer images. The HIRes 24 $\micron$ image presented in Fig.~\ref{fig:3a} \citep{Velusamy:14a} reaches a resolution of $\sim$ 2$\arcsec$; at this scale,  the extended atomic emission is directly comparable to the [FeII] and [NeII] line morphologies  seen in the SH maps of Fig.~\ref{fig:1}.  The HiRes IRAC image at 4.5~$\mu$m in Fig.~\ref{fig:3a}  at a resolution of $\sim$ 0.8~$\arcsec$ \citep{Velusamy:14a}  is dominated by highly excited H$_2$ and continuum emission from more evolved protostars. The HiRes technique reveals a number of structures which are not discernible in the IRAC image presented in Fig.~\ref{fig:1}: SMM1-b is coincident with bright spot of emission, and the near-IR source EC-41 \citep[e.g.][]{Hodapp:99a} becomes apparent at $\sim$~5$\arcsec$ to the north. A secondary bright spot to the NW of SMM1-b is likely related to  the 6.9~mm knot-E in \citet{Choi:09a}. The high-resolving power reached by the HiRes deconvolution method reveals emission structures in the Spitzer images that are in excellent agreement to the ones traced in the IRS line maps and the interferometric observations of \citet{Choi:09a}.

The emission morphology traced by the [FeII] and [NeII] lines, and the 24 $\micron$ HiRes image has the characteristics of a jet. To the resolution available, the emission is well aligned and confined in a single row of spaxels; it is detected very close to the protostellar sources, and it is terminated at the emission maxima. These positions likely correspond to terminating bow-shocks, as commonly observed in protostellar jets \citep[e.g., HH211,][]{Dionatos:10a}. However, the origin of the emission is not clear. It may represent either intrinsic atomic ejecta, as observed in evolved  young stellar objects, or emission produced in small shocks along the jet propagation axis, or both.

The morphologies of the atomic jet and the H$_2$ emission suggest that the former represents the intrinsic ejecta from the embedded protostar, SMM1, carving out a cavity on the protostellar envelope, which is delineated by the H$_2$ emission. The non-detected, symmetric cavity walls to the south in the NW lobe, could be attributed to a steep gradient in the local density in the north -- south direction. Such a description is compatible with the interpretation of the HCO$^+$ and HCN $J = 1-0$ interferometric maps of  \citet{Hogerheijde:99a}. 
Offsets between the atomic and H$_2$ emission have also been observed in HH54 \citep{Neufeld:06a}, as the two sets of lines trace very different excitation conditions in shocks.  However, the offsets observed in HH54 lie \textit{along} the propagation axis of the protostellar jet. 

In the present data the excitation of H$_2$ and the atomic lines occurs on two different axes and the molecular and atomic components are not co-spatial. Given that the pairs of atomic and H$_2$ lines presented in Fig.~\ref{fig:1} are separated by $\sim$~1~$\micron$ or less in wavelength, each pair of lines should be equally affected by extinction. The fact that the atomic jet is traced down to the protostellar source suggests it is less extinct than the H$_2$ emission. Close to embedded protostars, mid-IR emission is often highly extinct due to the dense envelope \citep[e.g.,][]{Dionatos:09a}, so the atomic emission likely originates from a region outside of the envelope and in the foreground of the Class~0 protostar SMM1-a.  Given the different position angles and extents of the low-$J$ H$_2$ and atomic components, the companion source SMM1-b suggested by \citet{Choi:09a} is a strong candidate as the driving source. It is well aligned with the atomic jet, and as a more evolved source it is expected to drive an atomic jet, as discussed in Sect.~\ref{sec:4}. SMM1-b shows strong emission in the deconvolved HiResIRAC image Fig.~\ref{fig:3a} in support of its classification as an evolved protostar.
  The radio jet  \citep{Choi:09a, Curiel:93a} and maser emission \citep{vanKempen:09a} associated with SMM1 extend at a position angle of $\sim$~135$^{\circ}$  in the same direction as the atomic jet. \citet{Choi:09a} associate the radio jet with the source SMM1-a and the H$_2$ emission with SMM1-b, despite that in their high resolution maps, SMM1-b lies on the radio jet axis and SMM1-a is offset by $\sim$~1$\arcsec$ to the NE from it. The morphological evidence presented here and the analysis in the following paragraphs suggest the opposite;  SMM1-a is responsible for the low-$J$ H$_2$ emission, and SMM1-b for the atomic and radio jets.  


\begin{figure*}
\centering
\resizebox{\hsize}{!}{\includegraphics{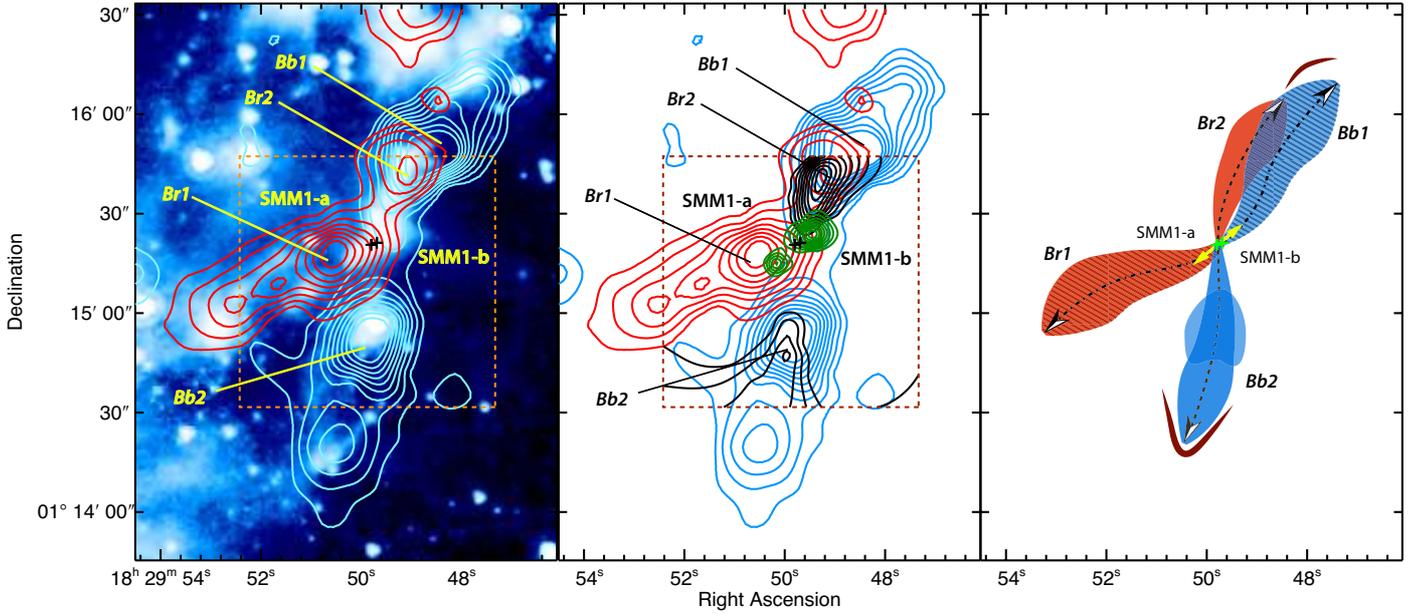}}
\caption{(\textit{Left:}) High velocity CO $J=3-2$ blue- and red-shifted emission (integrated over v$_{lsr}$ ranges of [-30, 0]  km s$^{-1}$ and [18, 48] km s$^{-1}$, rest v$_{lsr}$ = 8.8 km s$^{-1}$, and indicated as light blue and red solid contours, respectively) around SMM1 superimposed on an IRAC band 2 image. Outflow components are indicated using the nomenclature of \citet{Dionatos:10b} along with the positions of protostellar sources associated with the outflows. The dashed box delineates the limits of the LH \textit{Spitzer} maps. (\textit{Center:}) CO $J = 3-2$ emission as in the left panel, along with the H$_2$ S(1) (black) and [FeII] 18 $\micron$ (green) line maps. Notice the dimensions of the image in comparison to the details presented in Fig.~\ref{fig:1} and Fig.~\ref{fig:2}. H$_2$ emission follows well the pattern of CO lobes Br2 and Bb2 in a N-S orientation, whereas the [FeII] emission, despite its shorter extent, has the same orientation as the Bb1 - Br1 CO lobes pointing to the SE - NW direction. (\textit{Right:}) Sketch of the proposed outflow structure from SMM1-a and b. The outflow driven by the embedded source (SMM1-a) has an S-shaped pattern extending roughly in the N-S axis indicated with a dashed line and solid colors. The same outflow is most likely associated with the bow-shocks visible in the \textit{Spitzer} image of the left panel, delineated here as dark red V-shaped structures. The butterfly-like outflow extending in the SE-NW direction (delineated with a dashed-dotted line and hatched colors) most likely corresponds to the less embedded, foreground protostar (SMM1-b). The yellow arrows at the base of the outflows show the direction of the atomic jet.}
\label{fig:4}
\end{figure*}

In order to further examine the H$_2$ and atomic emission morphology, we employ the CO $J = 3 - 2$ data of the SMM1 region from \citet{Dionatos:10b} observed with JCMT. In the left panel of Fig.~\ref{fig:4}, we present the high velocity blue- and red-shifted CO emission, overlaid on the IRAC band-2 image. The CO maps display a complex morphology, especially towards the NW, where entangled blue- and red-shifted emission is seen extending in the same direction. The red-shifted lobe, Br2,  \citep[following the nomenclature of][]{Dionatos:10b}, is weaker and peaks closer to SMM1 than the blue-shifted lobe, Bb1. Strong red-shifted emission extends to the SE (lobe Br1) and blue-shifted gas directly to the south (Bb2). As noted in \citet{Dionatos:10b}, the pattern of the CO line-wings for Bb1 and Br1 shows signatures of high velocity "bullets", which are not observed towards the peaks of Bb2 and Br2. This characteristic provides a clear association between lobes Bb1 and  Br1 in a bipolar outflow scheme, and differentiates them from the Bb2 - Br2 pair.

The middle panel of Fig.\ref{fig:4} presents the 17~$\micron$ H$_2$ and the 18$\micron$ [FeII] line maps along with the CO outflow emission. It becomes apparent that the  H$_2$ emission is coincident with the CO lobes, Bb2 and Br2, in an almost north-south orientation. [FeII] emission is located at the base of the outflow complex and has practically the same orientation as the bipolar outflow Bb1- Br1, extending from the NW to the SE. The strongest atomic peak  at the NW is associated with the blue-shifted CO lobe Bb1 and the weaker atomic lobe observed at the SE with the red-shifted lobe Br1. The latter lobe is moving inwards the cloud to regions of higher extinction, which is consistent with the weaker atomic emission to the SE. The bright ridge to the NW seen in the IRAC image corresponds to the superposition of the Br2 and Bb1 outflows, which are driven by two different protostellar sources.

The right panel of  Fig.~\ref{fig:4} shows a sketch of the proposed scenario. The red-shifted emission to the north (Br2) is associated with the lobe Bb2 to the south (solid color lobes, dashed line), as indicated by the H$_2$ emission, and most likely is driven by the embedded protostellar source SMM1-a. The bipolar outflow Bb1 - Br1 (hatched color lobes, dot-dashed line) exhibiting ``bullet'' structure in CO, is associated with the atomic jet and most likely is driven by the less embedded protostar SMM1-b which lies in the foreground.  The binary source scenario can therefore disentangle the complex outflow morphology observed in CO, and is further supported by the excitation and dynamical properties of the [FeII] lines discussed in the next section.

\begin{table*}[!ht]
\caption{Line intensities at the peaks of emission, extracted for a region equal to the LH spaxel size}
\label{tab:1}
\centering
\begin{tabular}{l c r c c c }
\hline\hline
Element  &  Transition & E$_u$ (K) & $\lambda$ ($\mu$m)   &      \multicolumn{2}{c}{Intensity
(10 $^{-12}$ W cm$^{-2}$ sr$^{-1}$)} \\
 &   &  &  & blue lobe &  red lobe   \\

\hline
\hline

H$_2$  & $0 - 0$   S(2)    & 1681.76 &   12.2786   &   0.51$\pm$0.08  & $\ldots$   \\
$[$NeII$]$ & $^2$P$_{\frac{1}{2}}$ -- $^2$P$_{\frac{3}{2}}$ & 1122.85 &12.8135 & 2.09$\pm$0.09 & 0.47$\pm$0.09\\
H$_2$  & $0 - 0$   S(1)    & 1015.20 & 17.0348   &   0.41$\pm$0.07  & $\ldots$   \\
$[$FeII$]$ & $^4$F$_{\frac{7}{2}}$ -- $^4$F$_{\frac{9}{2}}$ & 3496.35 &17.9359 & 2.82$\pm$0.12 & 1.24$\pm$0.08\\
$[$FeII$]$ & $^4$F$_{\frac{5}{2}}$ -- $^4$F$_{\frac{7}{2}}$ & 4083.16 &24.5193 & 3.86$\pm$0.22 & $\ldots$ \\
$[$SI$]$ & $^3$P$_{1}$ -- $^3$P$_{2}$ &569.83 & 25.2490  & 5.28$\pm$0.26 & 1.92$\pm$0.24\\
$[$FeII$]$ & $^6$D$_{\frac{7}{2}}$ -- $^6$D$_{\frac{9}{2}}$ & 553.62 & 25.9883 & 11.02$\pm$0.37 & 8.22$\pm$0.18\\
$[$SiII$]$ & $^2$P$_{\frac{3}{2}}$ -- $^2$P$_{\frac{1}{2}}$ & 413.27 &34.8152 & 12.16$\pm$0.67 & 9.48$\pm$0.46\\
$[$FeII$]$ & $^6$D$_{\frac{5}{2}}$ -- $^6$D$_{\frac{7}{2}}$ & 960.64 &35.3487 & 3.34$\pm$0.47 & 2.31$\pm$0.38\\

\hline
\end{tabular}
\end{table*}

\section{Analysis and Discussion}\label{sec:4}

In order to assess the origin of the atomic emission and its possible progenitor, in the following sections we attempt to constrain its excitation conditions following both Local Thermodynamic Equilibrium (LTE) and non-LTE approaches. 

\subsection{LTE analysis}

Excitation diagrams are representations of the column density normalized for the upper level degeneracy of a transition and plotted against the corresponding upper level energy of that transition. Such diagrams provide a simple but powerful tool for examining the excitation of gas, assuming optically thin emission and LTE conditions. As long as these assumptions are valid, excitation diagrams can readily provide the excitation temperature and column density of  the species under consideration \citep{Goldsmith:99a}.

The excitation diagram method is commonly used for the analysis of molecular emission \citep[e.g.,][]{Neufeld:09a}, and is employed here for the study of atomic lines. The [FeII] lines arise from forbidden transitions with Einstein coefficients for spontaneous de-excitation in the order of 10$^{-30}$ s$^{-1}$, and thus are optically thin. Radiative rates, level energies and degeneracies for [FeII] were retrieved from \citet{Fuhr:07a} using the Atomic Spectra Database of the National Institute of Standards and Technology\footnote{http://www.nist.gov/index.html} \citep[NIST- ASD, ][]{Ralchenko:11a}.

\begin{figure}
\centering
\resizebox{\hsize}{!}{\includegraphics{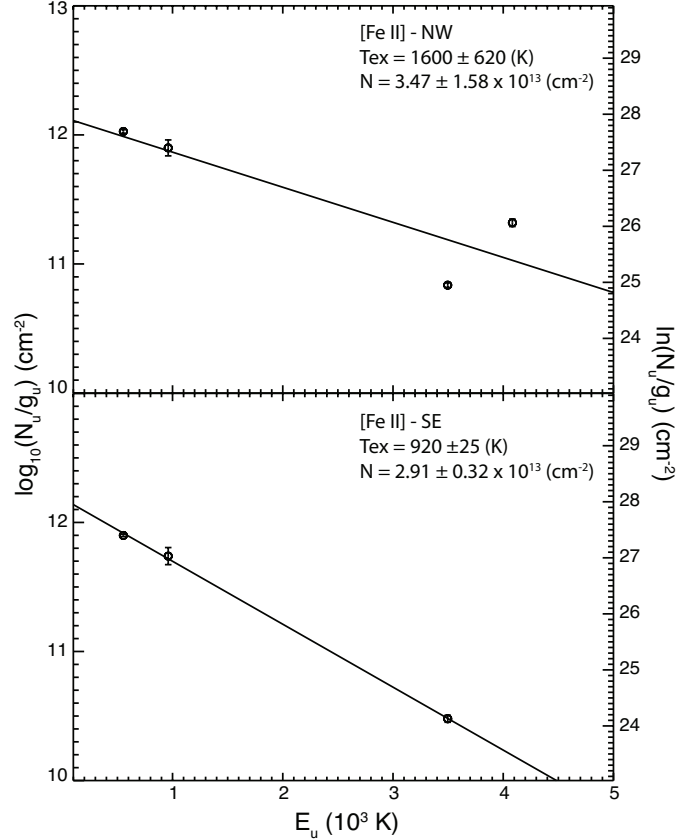}}
\caption{[FeII] excitation diagram for the NW and SE outflow lobes (upper and lower panels, respectively). The 24.5~$\micron$ line intensity at $E_u \sim$~4000~K is likely affected by residual bad pixel contamination and is thus excluded from the analysis. Derived temperatures and column densities are reported on the upper right corner of each panel.}
\label{fig:5}
\end{figure}

In Fig. \ref{fig:5},  we present the excitation diagrams for the NW and SE peaks of [FeII], sampled at the LH pixel scale (see Table~\ref{tab:1}). From the slope of the fitted lines, derived temperatures range between 1000 - 1600K, with the higher values being observed towards the NW. The slope of the fit at the NW position is strongly affected by the 24.5 $\micron$ line intensity (at $E_u \sim$ 4000 K), which may be enhanced due to residual bad pixel contamination. If ignored, the temperature for the two positions is consistent at $\sim$ 1000K. In excitation diagrams, column densities can be estimated from the intercept of the fitted line with the ordinate, given that the partition function at a given temperature is known. For the column density calculations in this work, we have employed the partition function values from NIST-ASD for the estimated range of temperatures. Derived column densities are $\sim 3 \times 10^{13}$ cm$^{-2}$ for the area covered by the LH spaxel. However the emitting area of [FeII] is likely smaller, so the present values are likely beam diluted and therefore represent lower limits. 

\subsection{non-LTE} 

For the non-LTE analysis of the [FeII] emission we employed the statistical equilibrium radiative transfer code, RADEX \citep{vanderTak:07a}. As in the case of LTE, we retrieved radiative rates, level energies and degeneracies from NIST-ASD. Having a first estimate of the excitation conditions from the LTE analysis, we have considered the excitation of iron through collisions with electrons. Collisions with atomic hydrogen become important only at much lower temperatures and densities \citep[see][]{Dionatos:09a}. Electron collisional rates were retrieved from the TIPbase of the IRON project\footnote{http://www.usm.uni-muenchen.de/people/ip/iron-project.html}, based on the calculations of \citet{Zhang:95a}.

\begin{figure}
\centering
\resizebox{\hsize}{!}{\includegraphics{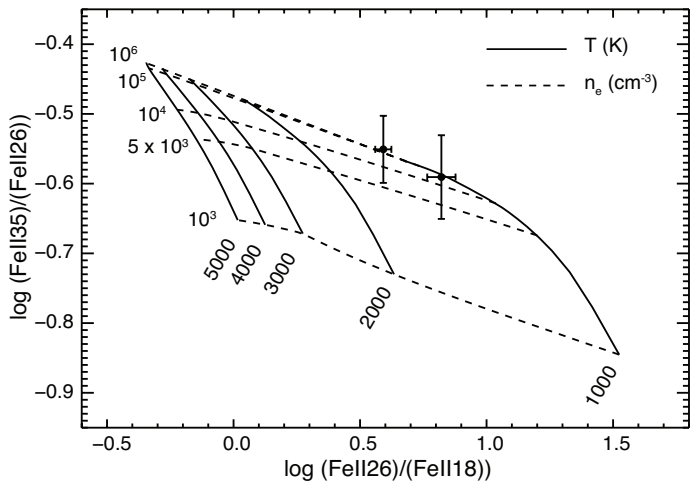}}
\caption{Non-LTE diagnostic diagram of the [FeII] emission for collisions with electrons created with RADEX. The temperature-sensitive 26/18~$\micron$ ratio (solid lines) is plotted against the electron density-sensitive 35/26~$\micron$ ratio (dashed lines). The observed ratios (filled circles with error-bars) indicate T $\sim$~1000 K  and n$_e$~$\sim$ 10$^5 - 10^6$~cm$^{-3}$.}
\label{fig:6}
\end{figure}

We ran RADEX for a grid of temperatures ranging from 1000 to 5000 K and densities between 10$^3$ and 10$^6$ cm $^{-3}$. The grid results are presented in Fig.~\ref{fig:6} in the form of line ratios. We employ the 26~$\micron$ over the 18~$\micron$ ratio as a temperature probe, since these lines arise from levels that are well separated in excitation, so that their population depends mainly on the efficiency of the excitation process (collisions with electrons in the current case). In addition the ratio of the 34.5~$\micron$ over the 26~$\micron$ lines is sensitive to density, as these lines arise from levels close in excitation energy but have different Einstein coefficients for spontaneous emission. The  24.5~$\micron$ to 18~$\micron$  ratio could also be used as a density probe, however the 24.5~$\micron$ line is detected at a single position and the LTE analysis has shown that it may be unreliable.

The observed line ratios in Fig.~\ref{fig:6} lie at temperatures of 1000-1200 K. This is in excellent agreement to the temperatures derived from the LTE analysis. The electron densities derived lie between $10^{5}$ and $10^{6}$ cm$^{-3}$, even though values as low as $5 \times 10^{3}$ are implied due to the high uncertainties of the ratio associated mostly to the uncertainties of the 34.5 $\micron$ line flux levels.

\subsection{Excitation of the atomic gas} 

Derived temperatures at the atomic jet peaks are consistent for both the LTE and non-LTE analysis, and lie at about $\sim$1000 K. This corresponds to ``hot'' gas for embedded sources, as derived in the mid-IR mainly from H$_2$ emission \citep{Lahuis:10a}. The H$_2$ emission in the line maps observed at the atomic peaks may be generated by the same processes, and the higher $J-$ H$_2$ transitions in the HiRes IRAC images show that the atomic and H$_2$ components are spatially coincident. However, the lack of higher $J-$ H$_2$ line maps does not allow us to directly compare the excitation conditions between the atomic and the hot H$_2$ components. In a broader sample of 43 embedded sources observed with \textit{Spitzer}/IRS \citep{Lahuis:10a}, only four display extended [FeII] 18~$\micron$ emission, indicative of highly excited atomic gas. Interestingly, two of them (SMM3 and SMM4) are also located in the Serpens molecular cloud.  For all these sources, a hot H$_2$ component is also present.\footnote{\citet{Lahuis:10a} report no hot component for SMM3 and SMM4 in Serpens, however this is well defined in \citep{Dionatos:13a}.}  

Electron densities between 10$^{5}$ and 10$^{6}$~cm$^{-3}$ are well above the levels of 500 - 1000~cm$^{-3}$ estimated in the case of L1448-mm, where no 18~$\micron$ [FeII] lines were detected \citep{Dionatos:09a}. The electron densities derived from the non-LTE analysis are consistent with the values typically observed in ``atomic" jets from Class~I protostars \citep{Nisini:02a}. However, typical temperatures for ``atomic" jets range between 7000 and 15000~K \citep{Takami:04a} which is much higher than the $\sim$~1000~K derived by our analysis. The near-IR [FeII] lines however, arise from transitions of much higher energies, and may as well correspond to a more energetic component of the jet or the shocks along its axis of propagation. Narrow band imaging for the typical ``atomic'' jet tracer [SII] $\lambda\lambda$~6716,~6731~\AA $ $  \citep{Davis:99a} does not trace any optical jet counterpart, indicating that the atomic jet observed here stands behind a layer of dust that obscures it in the visual bands. This is also suggested by the detection of extended CO$_2$ absorption bands seen also in the outflow positions (Fig.~\ref{fig:2}), and the extended continuum ridge lying across the atomic jet propagation axis as seen in the interferometric maps of \citet{Enoch:09a}.   The same absorbing layer is further obscuring also SMM1-a which lies in the background of SMM1-b (Sect.~\ref{sec:3}). 

The positioning of both sources behind the absorbing layer suggests that they are likely lying at similar distances from the observer. In addition, their angular separation of $\sim 1.5 \arcsec$ indicates that they are gravitationally bound forming a proto-binary system. Concerning the evolutionary stage of the sources, it is well established that SMM1-a is a deeply embedded Class 0 protostar \citep[e.g.][]{Larsson:00a}. This is also supported by the HiRes images of \citet{Velusamy:14a},  showing emission at 24~$\micron$ but not at 4.5~$\micron$. In stark contrast, SMM1-b is coincident with a bright point source at 4.5~$\micron$, which is in support of its classification by \citet{Choi:09a} as an evolved source. In conclusion, the current observational data suggest that the SMM1-a/SMM1-b system is a non-coeval proto-binary.  

Discussing the outflow morphology in Sect.~\ref{sec:3}, we concluded that the atomic jet is most likely associated with the high velocity "bullet" CO gas, reaching radial velocities as high as 50 km s$^{-1}$. \citet{Yildiz:13a} has demonstrated that the velocity traced by low-$J$ CO lines is only a lower limit as higher-$J$ transitions show increasingly higher velocity wings. The high velocity of the material associated with the atomic emission was also suggested by \citet{Goicoechea:12a}, who found the [OI] line at 63~$\micron$ shifted by 100~km~s$^{-1}$. Such high velocities, combined with the high densities derived from the non-LTE analysis indicate that at least part of the atomic emission is generated in dissociative shocks. Indeed, J-shock models predict that the [NeII] intensities observed require shock velocities higher than 70 km s$^{-1}$ \citep[][]{Hollenbach:89a}. J-shocks are also inferred by water maser emission  183 GHz along the atomic jet axis to the NW \citep[][see also Fig.~\ref{fig:1}]{vanKempen:09a}. As pointed out in \citet{Hollenbach:13a}, substantial maser emission in submillimeter wavelengths  requires temperatures $\gtrsim$~1000~K, which are readily produced behind J-shocks. Such temperatures are consistent with those derived from the [FeII] analysis.  
   
\subsection{On the driving source of the atomic jet}
 
The evolutionary status of a protostellar source may be inferred by the mass flux of the ejecta (see Sect.~\ref{sec:1}). For atomic emission, the mass flux of a jet can be derived using the method described in \citet{Dionatos:09a}. Summarizing, the method is based on the fact that the forbidden atomic line emission is optically thin and therefore the observed luminosity is proportional to the mass of the emitting gas. It requires knowledge of the velocity of the jet and relies on the assumption that no iron is locked onto dust grains. In the case of L1448-mm, however, which has similar high velocity``bullet'' gas as SMM1 \citep{Bachiller:90a, Kristensen:11a}, it has been shown that only a fraction between 5\% and 20\% of iron in the outflows is in the gas phase \citep{Dionatos:09a}.  Here, for the velocity of the atomic jet we adopt a conservative value of 100 km s$^{-1}$ \citep{Goicoechea:12a} and employ the iron gas phase fraction estimated in the case of L1448. Based on these assumptions, we find that the two-sided mass flux for the iron is $\sim 2 - 4 \times 10^{-7}$~M$_{\odot}$~yr$^{-1}$, consistent for all iron transitions.  For the adopted velocity, these values correspond to a jet momentum flux of $\sim$5$\times 10^{-5}$~M$_{\odot}$~km~s$^{-1}$~yr$^{-1}$, in very good agreement with the values derived for the CO lobes Bb1 and Br1 \citep{Dionatos:10b}. Despite the uncertainties in the derivation of the atomic jet momentum flux here, and the CO outflow momentum flux \citep[see][for extensive discussions]{Downes:07a, van_der_Marel:13a}, the estimations above suggest that the  [FeII] jet has enough thrust to power the large scale CO outflow.  



[NeII] emission at 12.8~$\micron$ may be produced through X-ray and FUV irradiation on outflow cavity walls and disk surfaces, or within shocks in protostellar jets. Examining a large sample of Class~II protostars, \citet{Guedel:10a} demonstrated that sources with jets have 1-2 orders of magnitude higher [NeII] luminosities compared to sources without jets. The same authors found that the mass loss rate in sources with jets closely correlates with the [NeII] line luminosity. This is also reflected in a decrease by $\sim$3 orders of magnitude in the [NeII] luminosity between Class~I and Class~III sources observed in the $\rho$ Oph cloud \citep{Flaccomio:09a}. The [NeII] emission detected around SMM1 is clearly extended and closely follows the pattern of the [FeII] jet, so there is little doubt on its origin. The [NeII] luminosity at the NW peak ranges from $\sim 9 \times 10^{28} - 2 \times 10^{29}$~erg~s$^{-1}$ for corresponding distances of 260 and 415 pc.  When compared to  the sample of YSO's in $\rho$-Oph \citep{Flaccomio:09a}, [NeII] luminosities lie at the border between Class~I and Class~II sources. For the estimated [NeII] luminosities, the X-wind model calculations of \citet{Shang:10a} predict a mass-loss rate of 10$^{-7.5} - 10^{-7}$~M$_{\odot}$~yr$^{-1}$,  in agreement with the values derived from [FeII]. 

In conclusion, the momentum flux derived by both [FeII] and [NeII] lines is $\sim$~10$^{-5}$~M$_{\odot}$~km~s$^{-1}$~yr$^{-1}$, assuming a jet velocity of 100 km s$^{-1}$. Based on the correlation between the momentum flux or force of the outflow and the bolometric luminosity of the source (F$_{CO}$/L$_{bol}$ ) of \citet{Cabrit:92a} for Class~0 protostars, SMM1-a with L$_{bol}$=71 L$_{\odot}$ \citep{Larsson:00a} would be expected to produce an CO outflow momentum flux of of 7 $\times 10^{-4}$~M$_{\odot}$~km~s$^{-1}$~yr$^{-1}$, almost two orders of magnitude higher than the values estimated here.  Consequently, the values here are low for a deeply embedded Class~0 protostar, and are compatible with a rather evolved Class~I/II source \citep{Hartigan:94a, Cabrit:07a}.  This is in line with the characterization of SMM1-b as a disk source from the SED slope between 7~mm and 6.9~cm \citep{Choi:09a}, and provides additional evidence that the atomic jet is indeed driven by the more evolved companion source to the Class~0 protostar SMM1-a.

\section{Conclusions}\label{sec:5}

We have carried out spectro-imaging observations of the region around SMM1 in Serpens with \textit{Spitzer}/IRS, encompassing a wavelength range between 10 and 38~$\micron$. These observations trace atomic ([FeII], [NeII], [SI] and [SiII]) and molecular (H$_2$) emission extending from the protostellar source. \textit{We have in addition compared the emission line maps to deconvolved HiRes IRAC and MIPS images}  The main results are summarized as follows:

\begin{itemize}

\item{The position angles of low-$J$ H$_2$ and atomic emission differ by $\sim$~35$^\circ$. Atomic emission is traced at distances $< 5 \arcsec$ while H$_2$ becomes prominent at larger distances from SMM1. Given that pairs of H$_2$ and atomic lines are separated by $\sim$~1$\micron$ or less, the atomic emission is less extinct and lies in front of the dense envelope surrounding the embedded source SMM1-a.  Therefore the atomic emission originates from a companion source lying in the foreground.}

\item{The H$_2$ and atomic emission disentangle the large-scale CO outflow structure into two outflows. The H$_2$ emission corresponds to the CO outflow extending roughly in the N-S direction, while the atomic emission drives the high velocity CO outflow extending in the NW-SE direction. The disentangled outflow morphology is compatible only with a proto-binary source. }

\item{LTE and non-LTE analysis of the [FeII] lines suggest an excitation temperature of $\sim$~1000K. Electron densities  between 10$^{-5}$ and 10$^{-6}$~cm$^{-3}$ are consistent with the values observed in atomic jets from evolved protostars. However the temperatures traced by higher excitation [FeII] lines in the near-IR are an order of magnitude higher than the ones traced here. }

\item{The atomic jet has not been traced before in visual or near-IR wavelengths, as it lies behind a dust layer revealed by continuum emission and CO$_2$ absorption bands.}

\item{The mass flux estimated from the [FeII] ranges between $2 - 4 \times 10^{-7}$~M${_\odot}$~yr$^{-1}$ for both sides of the atomic jet.  This corresponds to a momentum flux of $\sim5 \times 10^{-5}$~M$_{\odot}$~km~s$^{-1}$~yr$^{-1}$, which is in agreement with the values corresponding to the NW-SE CO outflow. Thus the atomic jet has enough thrust to drive the large scale CO outflow. }

\item{Compared with X-wind models, the observed [NeII] line luminosity is consistent with mass-loss rates between $10^{-7}$ and $10^{-7.5}$ M$_{\odot}$ yr$^{-1}$.    }

\item{Mass-loss rates from both tracers are compatible with measurements for rather evolved Class~I/II sources.}

\item{The luminosity of the [NeII] lines compared with the luminosities of protostellar sources in the $\rho$-Oph cloud is also in agreement  that the driving source of the atomic jet is a Class~I/II protostar.}

\end{itemize} 

The existence of a less embedded companion source to the embedded protostellar source SMM1 has long been suspected \citep[e.g.][]{Eiroa:89a, Hodapp:99a, Eiroa:05a, Choi:09a, Dionatos:10b}.  The atomic lines detected in this work reveal  for the first time the ejecta from the binary companion.  Our analysis gives  ample evidence that the atomic jet is compatible with a more evolved Class~I/II source.  Most likely, the companion source driving the atomic jet is SMM1-b. We therefore conclude that SMM1 is a non-coeval proto-binary source, consisting of the known embedded Class~0 protostar SMM1-a (SMM1/FIRS1) and a more evolved Class~I/II companion SMM1-b, lying at $\sim 1.5\arcsec$ to the NW and in the foreground of SMM1-a.

\begin{acknowledgements}

We thank the referee (T. Velusamy) for his very constructive comments and for providing the Spitzer HiRes images presented here. This research was supported by a grant from the Instrument Center for Danish Astrophysics (IDA), a Lundbeck Foundation Group Leader Fellowship to JKJ and from the EU FP7-2011 under Grant Agreement nr. 284405. This publication is supported by the Austrian Science Fund (FWF). Research at Centre for Star and Planet Formation is funded by the Danish National Research Foundation and the University of Copenhagen's programme of excellence.

 \end{acknowledgements}

\begin{tiny}

\bibliographystyle{aa}
\bibliography{smm1}

\begin{thebibliography}{61}
\expandafter\ifx\csname natexlab\endcsname\relax\def\natexlab#1{#1}\fi

\bibitem[{{Agra-Amboage} {et~al.}(2011){Agra-Amboage}, {Dougados}, {Cabrit}, \&
  {Reunanen}}]{Agra-Amboage:11a}
{Agra-Amboage}, V., {Dougados}, C., {Cabrit}, S., \& {Reunanen}, J. 2011, \aap,
  532, A59

\bibitem[{{Arce} \& {Sargent}(2006)}]{Arce:06a}
{Arce}, H.~G. \& {Sargent}, A.~I. 2006, \apj, 646, 1070

\bibitem[{{Bachiller} {et~al.}(1990){Bachiller}, {Martin-Pintado}, {Tafalla},
  {Cernicharo}, \& {Lazareff}}]{Bachiller:90a}
{Bachiller}, R., {Martin-Pintado}, J., {Tafalla}, M., {Cernicharo}, J., \&
  {Lazareff}, B. 1990, \aap, 231, 174

\bibitem[{{Bontemps} {et~al.}(1996){Bontemps}, {Andre}, {Terebey}, \&
  {Cabrit}}]{Bontemps:96a}
{Bontemps}, S., {Andre}, P., {Terebey}, S., \& {Cabrit}, S. 1996, \aap, 311,
  858

\bibitem[{{Cabrit}(2007)}]{Cabrit:07a}
{Cabrit}, S. 2007, in Lecture Notes in Physics, Berlin Springer Verlag, Vol.
  723, Lecture Notes in Physics, Berlin Springer Verlag, ed. J.~{Ferreira},
  C.~{Dougados}, \& E.~{Whelan}, 21

\bibitem[{{Cabrit} \& {Bertout}(1992)}]{Cabrit:92a}
{Cabrit}, S. \& {Bertout}, C. 1992, \aap, 261, 274

\bibitem[{{Choi}(2009)}]{Choi:09a}
{Choi}, M. 2009, \apj, 705, 1730

\bibitem[{{Curiel} {et~al.}(1993){Curiel}, {Rodriguez}, {Moran}, \&
  {Canto}}]{Curiel:93a}
{Curiel}, S., {Rodriguez}, L.~F., {Moran}, J.~M., \& {Canto}, J. 1993, \apj,
  415, 191

\bibitem[{{Davis} {et~al.}(1999){Davis}, {Matthews}, {Ray}, {Dent}, \&
  {Richer}}]{Davis:99a}
{Davis}, C.~J., {Matthews}, H.~E., {Ray}, T.~P., {Dent}, W.~R.~F., \& {Richer},
  J.~S. 1999, \mnras, 309, 141

\bibitem[{{Dionatos} {et~al.}(2013){Dionatos}, {J{\o}rgensen}, {Green},
  {Herczeg}, {Evans}, {Kristensen}, {Lindberg}, \& {van
  Dishoeck}}]{Dionatos:13a}
{Dionatos}, O., {J{\o}rgensen}, J.~K., {Green}, J.~D., {et~al.} 2013, ArXiv
  e-prints

\bibitem[{{Dionatos} {et~al.}(2010{\natexlab{a}}){Dionatos}, {Nisini},
  {Cabrit}, {Kristensen}, \& {Pineau Des For{\^e}ts}}]{Dionatos:10a}
{Dionatos}, O., {Nisini}, B., {Cabrit}, S., {Kristensen}, L., \& {Pineau Des
  For{\^e}ts}, G. 2010{\natexlab{a}}, \aap, 521, A7

\bibitem[{{Dionatos} {et~al.}(2010{\natexlab{b}}){Dionatos}, {Nisini},
  {Codella}, \& {Giannini}}]{Dionatos:10b}
{Dionatos}, O., {Nisini}, B., {Codella}, C., \& {Giannini}, T.
  2010{\natexlab{b}}, \aap, 523, A29

\bibitem[{{Dionatos} {et~al.}(2009){Dionatos}, {Nisini}, {Garcia Lopez},
  {Giannini}, {Davis}, {Smith}, {Ray}, \& {DeLuca}}]{Dionatos:09a}
{Dionatos}, O., {Nisini}, B., {Garcia Lopez}, R., {et~al.} 2009, \apj, 692, 1

\bibitem[{{Downes} \& {Cabrit}(2007)}]{Downes:07a}
{Downes}, T.~P. \& {Cabrit}, S. 2007, \aap, 471, 873

\bibitem[{{Dzib} {et~al.}(2010){Dzib}, {Loinard}, {Mioduszewski}, {Boden},
  {Rodr{\'{\i}}guez}, \& {Torres}}]{Dzib:10a}
{Dzib}, S., {Loinard}, L., {Mioduszewski}, A.~J., {et~al.} 2010, \apj, 718, 610

\bibitem[{{Eiroa} \& {Casali}(1989)}]{Eiroa:89a}
{Eiroa}, C. \& {Casali}, M.~M. 1989, \aap, 223, L17

\bibitem[{{Eiroa} {et~al.}(2005){Eiroa}, {Torrelles}, {Curiel}, \&
  {Djupvik}}]{Eiroa:05a}
{Eiroa}, C., {Torrelles}, J.~M., {Curiel}, S., \& {Djupvik}, A.~A. 2005, \aj,
  130, 643

\bibitem[{{Enoch} {et~al.}(2009){Enoch}, {Corder}, {Dunham}, \&
  {Duch{\^e}ne}}]{Enoch:09a}
{Enoch}, M.~L., {Corder}, S., {Dunham}, M.~M., \& {Duch{\^e}ne}, G. 2009, \apj,
  707, 103

\bibitem[{{Ferreira}(1997)}]{Ferreira:97a}
{Ferreira}, J. 1997, \aap, 319, 340

\bibitem[{{Flaccomio} {et~al.}(2009){Flaccomio}, {Stelzer}, {Sciortino},
  {Micela}, {Pillitteri}, \& {Testi}}]{Flaccomio:09a}
{Flaccomio}, E., {Stelzer}, B., {Sciortino}, S., {et~al.} 2009, \aap, 505, 695

\bibitem[{{Goicoechea} {et~al.}(2012){Goicoechea}, {Cernicharo}, {Karska},
  {Herczeg}, {Polehampton}, {Wampfler}, {Kristensen}, {van Dishoeck},
  {Etxaluze}, {Bern{\'e}}, \& {Visser}}]{Goicoechea:12a}
{Goicoechea}, J.~R., {Cernicharo}, J., {Karska}, A., {et~al.} 2012, \aap, 548,
  A77

\bibitem[{{Goldsmith} \& {Langer}(1999)}]{Goldsmith:99a}
{Goldsmith}, P.~F. \& {Langer}, W.~D. 1999, \apj, 517, 209

\bibitem[{{Green} {et~al.}(2013){Green}, {Evans}, {J{\o}rgensen}, {Herczeg},
  {Kristensen}, {Lee}, {Dionatos}, {Yildiz}, {Salyk}, {Meeus}, {Bouwman},
  {Visser}, {Bergin}, {van Dishoeck}, {Rascati}, {Karska}, {van Kempen},
  {Dunham}, {Lindberg}, {Fedele}, \& {DIGIT Team}}]{Green:13a}
{Green}, J.~D., {Evans}, II, N.~J., {J{\o}rgensen}, J.~K., {et~al.} 2013, \apj,
  770, 123

\bibitem[{{G{\"u}del} {et~al.}(2010){G{\"u}del}, {Lahuis}, {Briggs}, {Carr},
  {Glassgold}, {Henning}, {Najita}, {van Boekel}, \& {van
  Dishoeck}}]{Guedel:10a}
{G{\"u}del}, M., {Lahuis}, F., {Briggs}, K.~R., {et~al.} 2010, \aap, 519, A113

\bibitem[{{Gueth} \& {Guilloteau}(1999)}]{Gueth:99a}
{Gueth}, F. \& {Guilloteau}, S. 1999, \aap, 343, 571

\bibitem[{{Hartigan} {et~al.}(1995){Hartigan}, {Edwards}, \&
  {Ghandour}}]{Hartigan:95a}
{Hartigan}, P., {Edwards}, S., \& {Ghandour}, L. 1995, \apj, 452, 736

\bibitem[{{Hartigan} {et~al.}(1994){Hartigan}, {Morse}, \&
  {Raymond}}]{Hartigan:94a}
{Hartigan}, P., {Morse}, J.~A., \& {Raymond}, J. 1994, \apj, 436, 125

\bibitem[{{Hodapp}(1999)}]{Hodapp:99a}
{Hodapp}, K.~W. 1999, \aj, 118, 1338

\bibitem[{{Hogerheijde} {et~al.}(1999){Hogerheijde}, {van Dishoeck},
  {Salverda}, \& {Blake}}]{Hogerheijde:99a}
{Hogerheijde}, M.~R., {van Dishoeck}, E.~F., {Salverda}, J.~M., \& {Blake},
  G.~A. 1999, \apj, 513, 350

\bibitem[{{Hollenbach} {et~al.}(2013){Hollenbach}, {Elitzur}, \&
  {McKee}}]{Hollenbach:13a}
{Hollenbach}, D., {Elitzur}, M., \& {McKee}, C.~F. 2013, ArXiv e-prints

\bibitem[{{Hollenbach} \& {McKee}(1989)}]{Hollenbach:89a}
{Hollenbach}, D. \& {McKee}, C.~F. 1989, \apj, 342, 306

\bibitem[{{Houck} {et~al.}(2004){Houck}, {Roellig}, {van Cleve}, {Forrest},
  {Herter}, {Lawrence}, {Matthews}, {Reitsema}, {Soifer}, {Watson}, {Weedman},
  {Huisjen}, {Troeltzsch}, {Barry}, {Bernard-Salas}, {Blacken}, {Brandl},
  {Charmandaris}, {Devost}, {Gull}, {Hall}, {Henderson}, {Higdon}, {Pirger},
  {Schoenwald}, {Sloan}, {Uchida}, {Appleton}, {Armus}, {Burgdorf},
  {Fajardo-Acosta}, {Grillmair}, {Ingalls}, {Morris}, \& {Teplitz}}]{Houck:04a}
{Houck}, J.~R., {Roellig}, T.~L., {van Cleve}, J., {et~al.} 2004, \apjs, 154,
  18

\bibitem[{{Kristensen} {et~al.}(2011){Kristensen}, {van Dishoeck}, {Tafalla},
  {Bachiller}, {Nisini}, {Liseau}, \& {Y{\i}ld{\i}z}}]{Kristensen:11a}
{Kristensen}, L.~E., {van Dishoeck}, E.~F., {Tafalla}, M., {et~al.} 2011, \aap,
  531, L1

\bibitem[{{Lahuis} {et~al.}(2010){Lahuis}, {van Dishoeck}, {J{\o}rgensen},
  {Blake}, \& {Evans}}]{Lahuis:10a}
{Lahuis}, F., {van Dishoeck}, E.~F., {J{\o}rgensen}, J.~K., {Blake}, G.~A., \&
  {Evans}, N.~J. 2010, \aap, 519, A3

\bibitem[{{Larsson} {et~al.}(2000){Larsson}, {Liseau}, {Men'shchikov},
  {Olofsson}, {Caux}, {Ceccarelli}, {Lorenzetti}, {Molinari}, {Nisini},
  {Nordh}, {Saraceno}, {Sibille}, {Spinoglio}, \& {White}}]{Larsson:00a}
{Larsson}, B., {Liseau}, R., {Men'shchikov}, A.~B., {et~al.} 2000, \aap, 363,
  253

\bibitem[{{Neufeld} {et~al.}(2006){Neufeld}, {Melnick}, {Sonnentrucker},
  {Bergin}, {Green}, {Kim}, {Watson}, {Forrest}, \& {Pipher}}]{Neufeld:06a}
{Neufeld}, D.~A., {Melnick}, G.~J., {Sonnentrucker}, P., {et~al.} 2006, \apj,
  649, 816

\bibitem[{{Neufeld} {et~al.}(2009){Neufeld}, {Nisini}, {Giannini}, {Melnick},
  {Bergin}, {Yuan}, {Maret}, {Tolls}, {G{\"u}sten}, \& {Kaufman}}]{Neufeld:09a}
{Neufeld}, D.~A., {Nisini}, B., {Giannini}, T., {et~al.} 2009, \apj, 706, 170

\bibitem[{{Nisini} {et~al.}(2002){Nisini}, {Caratti o Garatti}, {Giannini}, \&
  {Lorenzetti}}]{Nisini:02a}
{Nisini}, B., {Caratti o Garatti}, A., {Giannini}, T., \& {Lorenzetti}, D.
  2002, \aap, 393, 1035

\bibitem[{{Panoglou} {et~al.}(2012){Panoglou}, {Cabrit}, {Pineau Des
  For{\^e}ts}, {Garcia}, {Ferreira}, \& {Casse}}]{Panoglou:12a}
{Panoglou}, D., {Cabrit}, S., {Pineau Des For{\^e}ts}, G., {et~al.} 2012, \aap,
  538, A2

\bibitem[{{Ralchenko} {et~al.}(2011){Ralchenko}, {Kramida}, {Reader}, \& {the
  NIST ASD Team}}]{Ralchenko:11a}
{Ralchenko}, Y., {Kramida}, A.~E., {Reader}, J., \& {the NIST ASD Team}. 2011,
  Available: http://physics.nist.gov/asd, 309, 1

\bibitem[{{Shang} {et~al.}(2010){Shang}, {Glassgold}, {Lin}, \&
  {Liu}}]{Shang:10a}
{Shang}, H., {Glassgold}, A.~E., {Lin}, W.-C., \& {Liu}, C.-F.~J. 2010, \apj,
  714, 1733

\bibitem[{{Shu} {et~al.}(1994){Shu}, {Najita}, {Ostriker}, {Wilkin}, {Ruden},
  \& {Lizano}}]{Shu:94a}
{Shu}, F., {Najita}, J., {Ostriker}, E., {et~al.} 1994, \apj, 429, 781

\bibitem[{{Smith} {et~al.}(2007){Smith}, {Armus}, {Dale}, {Roussel}, {Sheth},
  {Buckalew}, {Jarrett}, {Helou}, \& {Kennicutt}}]{Smith:07a}
{Smith}, J.~D.~T., {Armus}, L., {Dale}, D.~A., {et~al.} 2007, \pasp, 119, 1133

\bibitem[{{Strai{\v z}ys} {et~al.}(2003){Strai{\v z}ys}, {{\v C}ernis}, \&
  {Barta{\v s}i{\= u}t{\.e}}}]{Straizys:03a}
{Strai{\v z}ys}, V., {{\v C}ernis}, K., \& {Barta{\v s}i{\= u}t{\.e}}, S. 2003,
  \aap, 405, 585

\bibitem[{{Takami} {et~al.}(2004){Takami}, {Chrysostomou}, {Ray}, {Davis},
  {Dent}, {Bailey}, {Tamura}, \& {Terada}}]{Takami:04a}
{Takami}, M., {Chrysostomou}, A., {Ray}, T.~P., {et~al.} 2004, \aap, 416, 213

\bibitem[{{Testi} \& {Sargent}(1998)}]{Testi:98a}
{Testi}, L. \& {Sargent}, A.~I. 1998, \apjl, 508, L91

\bibitem[{{van der Marel} {et~al.}(2013){van der Marel}, {Kristensen},
  {Visser}, {Mottram}, {Y{\i}ld{\i}z}, \& {van Dishoeck}}]{van_der_Marel:13a}
{van der Marel}, N., {Kristensen}, L.~E., {Visser}, R., {et~al.} 2013, ArXiv
  e-prints

\bibitem[{{van der Tak} {et~al.}(2007){van der Tak}, {Black}, {Sch{\"o}ier},
  {Jansen}, \& {van Dishoeck}}]{vanderTak:07a}
{van der Tak}, F.~F.~S., {Black}, J.~H., {Sch{\"o}ier}, F.~L., {Jansen}, D.~J.,
  \& {van Dishoeck}, E.~F. 2007, \aap, 468, 627

\bibitem[{{van Kempen} {et~al.}(2009){van Kempen}, {Wilner}, \&
  {Gurwell}}]{vanKempen:09a}
{van Kempen}, T.~A., {Wilner}, D., \& {Gurwell}, M. 2009, \apjl, 706, L22

\bibitem[{{Velusamy} {et~al.}(2011){Velusamy}, {Langer}, {Kumar}, \&
  {Grave}}]{Velusamy:11a}
{Velusamy}, T., {Langer}, W.~D., {Kumar}, M.~S.~N., \& {Grave}, J.~M.~C. 2011,
  \apj, 741, 60

\bibitem[{{Velusamy} {et~al.}(2007){Velusamy}, {Langer}, \&
  {Marsh}}]{Velusamy:07a}
{Velusamy}, T., {Langer}, W.~D., \& {Marsh}, K.~A. 2007, \apjl, 668, L159

\bibitem[{{Velusamy} {et~al.}(2013){Velusamy}, {Langer}, \&
  {Thompson}}]{Velusamy:14a}
{Velusamy}, T., {Langer}, W.~D., \& {Thompson}, T. 2013, ArXiv e-prints

\bibitem[{{Velusamy} {et~al.}(2008){Velusamy}, {Marsh}, {Beichman}, {Backus},
  \& {Thompson}}]{Velusamy:08a}
{Velusamy}, T., {Marsh}, K.~A., {Beichman}, C.~A., {Backus}, C.~R., \&
  {Thompson}, T.~J. 2008, \aj, 136, 197

\bibitem[{{Visser} {et~al.}(2012){Visser}, {Kristensen}, {Bruderer}, {van
  Dishoeck}, {Herczeg}, {Brinch}, {Doty}, {Harsono}, \& {Wolfire}}]{Visser:12a}
{Visser}, R., {Kristensen}, L.~E., {Bruderer}, S., {et~al.} 2012, \aap, 537,
  A55

\bibitem[{{White} {et~al.}(1995){White}, {Casali}, \& {Eiroa}}]{White:95a}
{White}, G.~J., {Casali}, M.~M., \& {Eiroa}, C. 1995, \aap, 298, 594

\bibitem[{{Wiese} \& {Fuhr}(2007)}]{Fuhr:07a}
{Wiese}, J.~R. \& {Fuhr}, W.~L. 2007, J. Phys. Chem. Ref. Data, 35, 1669

\bibitem[{{Winston} {et~al.}(2007){Winston}, {Megeath}, {Wolk}, {Muzerolle},
  {Gutermuth}, {Hora}, {Allen}, {Spitzbart}, {Myers}, \& {Fazio}}]{Winston:07a}
{Winston}, E., {Megeath}, S.~T., {Wolk}, S.~J., {et~al.} 2007, \apj, 669, 493

\bibitem[{{Wu} {et~al.}(2004){Wu}, {Wei}, {Zhao}, {Shi}, {Yu}, {Qin}, \&
  {Huang}}]{Wu:04a}
{Wu}, Y., {Wei}, Y., {Zhao}, M., {et~al.} 2004, \aap, 426, 503

\bibitem[{{Y{\i}ld{\i}z} {et~al.}(2013){Y{\i}ld{\i}z}, {Kristensen}, {van
  Dishoeck}, {San Jose-Garcia}, {Karska}, {Harsono}, {Tafalla}, {Fuente},
  {Visser}, {J{\o}rgensen}, \& {Hogerheijde}}]{Yildiz:13a}
{Y{\i}ld{\i}z}, U.~A., {Kristensen}, L.~E., {van Dishoeck}, E.~F., {et~al.}
  2013, ArXiv e-prints

\bibitem[{{Zhang} \& {Pradhan}(1995)}]{Zhang:95a}
{Zhang}, H.~L. \& {Pradhan}, A.~K. 1995, \aap, 293, 953

\bibitem[{{Zhang} {et~al.}(2005){Zhang}, {Hunter}, {Brand}, {Sridharan},
  {Cesaroni}, {Molinari}, {Wang}, \& {Kramer}}]{Zhang:05a}
{Zhang}, Q., {Hunter}, T.~R., {Brand}, J., {et~al.} 2005, \apj, 625, 864

\end{thebibliography}

\end{tiny}

\end{document}